\documentclass{Files/llncs}
\usepackage{Files/llncsdoc}
\usepackage{times}
\usepackage{helvet}
\usepackage{courier}
\usepackage{paralist}
\usepackage{amssymb}
\usepackage{graphicx}
\usepackage{epsfig}
\usepackage{sidecap}
\usepackage{wrapfig}
\usepackage{a4,a4wide}

\usepackage[vlined,boxed]{Files/algorithm2e}

\newcommand{\ignore}[1]{} 

\newcommand{\here}[1]{{\bf *** #1 ***}}

\newcommand{\cL}{{\cal L}}
\newcommand{\cU}{{\cal U}}
\newcommand{\cA}{{\cal A}}
\newcommand{\cB}{{\cal B}}

\begin{document}

\title{Solving Modular Model Expansion Tasks}
\author{Shahab Tasharrofi, Xiongnan (Newman) Wu, Eugenia Ternovska}
\institute{Simon Fraser University\\ \{sta44,xwa33,ter\}@cs.sfu.ca}
\maketitle

\begin{abstract}

The work we describe here is a part of a research program of developing
foundations of declarative solving of search problems.
We consider the model expansion
task as the task representing the essence of search problems where we are given
an instance of a problem and are searching for a solution satisfying
certain properties. Such tasks are common in artificial intelligence,
formal verification, computational biology.
Recently, the model expansion framework was extended to deal
with multiple modules. In the current paper, inspired by practical combined solvers,
we introduce an algorithm to solve model expansion tasks for modular systems.
We show that our
algorithm closely corresponds to what is done in practice in  different areas such as Satisfiability Modulo Theories (SMT),
Integer Linear Programming (ILP), Answer Set Programming (ASP).
\end{abstract}

\section{Introduction}
\label{sec:intro}

The  research described in this paper is a part of a research program
of developing formal foundations for specification/modelling languages
(declarative programming) for solving computationally hard  problems.
In \cite{MT05}, the authors  formalize search problems as the logical task 
of {\em model expansion (MX)}, the task of expanding a given (mathematical)
structure with new relations. They started a research program of finding
common underlying principles of various approaches to specifying and solving
search problems, finding appropriate mathematical abstractions, and
investigating complexity-theoretic and expressiveness issues. It was
emphasized that it is important to understand the expressiveness of a
specification language in terms of the computational complexity of the
problems it can represent. Complexity-theoretic aspects of model expansion
for several logics in the context of related computational tasks of
satisfiability and model checking were studied in \cite{Kolokolova:complexity:LPAR:long}.
Since built-in arithmetic is present in all realistic modelling languages,
it was important to formalize built-in arithmetic in such languages. In
\cite{TM:IJCAI:2009}, model expansion ideas were extended to provide
mathematical foundation for dealing with arithmetic and aggregate functions
($min$, $sum$ etc.). There, the instance and expansion structures are
embedded into an infinite structure of arithmetic, and the property of
capturing NP was proven for a logic which corresponds to practical
languages. The proposed formalism applies to other infinite background
structures besides arithmetic. The analysis of practical languages was given
in \cite{TT:NonMon30}. It was proved that certain common problems involving
numbers (e.g. integer factorization) are not expressible in the ASP and IDP
system languages naturally, and in \cite{TT:arithmetic:LPAR-17:long}, the authors
improved the result of \cite{TM:IJCAI:2009} by defining a new logic which
unconditionally captures NP over arithmetical structures.

The next step in the development of the MX-based framework is adding modularity
concepts. It is convenient from the point of view of a user to be able to
split a large problem into subproblems, and to use the most suitable formalism
for each part, and thus a unifying semantics is needed. In a recent work
\cite{TT:FROCOS:2011-long}, a subset of the authors extended the MX framework to be
able to represent a modular system. The most interesting aspect of that
proposal is that modules can be considered from both model-theoretic and
operational view. Under the model-theoretic view, an MX module is a set (or
class) of structures, and under the operational view it is an operator, mapping
a subset of the vocabulary to another subset. An abstract algebra on MX modules
is given, and it allows one to combine modules on abstract model-theoretic
level, independently from what languages are used for describing them. Perhaps
the most important operation in the algebra is the loop (or feedback)
operation, since iteration underlies many solving methods. The authors show
that the power of the loop operator is such that the combined modular system
can capture all of the complexity class NP even when each module is
deterministic and polytime. Moreover, in general, adding loops gives a jump in
the polynomial time hierarchy, one step from the highest complexity of the
components. It is also shown that each module can be viewed as an operator, and
when each module is (anti-) monotone, the number of the potential solutions can
be significantly reduced by using ideas from the logic programming community.

To develop the framework further, we need a method for ``solving'' modular MX
systems. By solving we mean finding structures which are in the modular system,
where the system is viewed as a function of individual modules. {\em Our goal
is to come up with a general algorithm which takes a modular system in input
and generates its solutions.}

We take our inspiration in how ``combined'' solvers are constructed in the
general field of declarative problem solving. The field consists of many areas
such as Integer Linear Programming (ILP), Answer Set Programming (ASP),
Satisfiability Modulo Theories (SMT), Satisfiability (SAT), and Constraint
Programming (CP), and each of these areas has developed multitudes of solvers,
including powerful ``combined'' solvers such as SMT solvers. Moreover, SMT-like
techniques are needed in the ASP community \cite{Niemel:LPNMR2009}. Our main
challenge is to come up with an appropriate mathematical abstraction of
``combined'' solving. Our contributions are as follows.
\begin{compactenum}
  \item We formalize common principles of ``combined'' solving in different
communities in the context of modular model expansion. Just as in
\cite{TT:FROCOS:2011-long}, we  use a combination of a model-theoretic, algebraic and
operational view of modular systems.

  \item We design an abstract algorithm that given a modular system, computes the
models of that modular system iteratively, and we formulate conditions on
languages of individual modules to participate in the iterative solving. We use
the formalization above of these common principles to show the effectiveness of
our algorithm.

  \item We introduce abstractions for many ideas in practical systems such as the
concept of a {\em valid acceptance procedure} that abstractly represents unit
propagation in SAT, well-founded model computation in ASP, arc-consistency
checkers in CP, etc.

  \item As a proof of concept, we show that, in the context of the model
expansion task, our algorithm generalizes the work of different solvers from
different communities in a unifying and abstract way. In particular, we
investigate the branch-and-cut technique in ILP and methods used in SMT, DPLL(Agg)
and combinations of ASP and CP \cite{ILP,SMT,DPLL_AGG,ASP-CP-combination}. We aim
to show that, although no implementation is presented, the algorithm should work
fine as it mimics the current technology.

  \item We develop an improvement of our algorithm by using approximation methods
proposed in \cite{TT:FROCOS:2011-long}.
\end{compactenum}


\section{Background}
\label{sec:background}

\subsection{Model Expansion} In  \cite{MT05}, the authors  formalize combinatorial search problems  
 as the  task of
{\em model expansion (MX)}, the logical task of expanding a given (mathematical)
structure with new relations. 
Formally, the user axiomatizes the problem in
some logic $\cal L$. This axiomatization relates an instance of the problem (a
{\em finite structure}, i.e., a universe together with some relations and functions),
and its solutions (certain {\em expansions} of that structure with new relations or
functions). Logic $\cal L$ corresponds to a specification/modelling language. It
could be an extension of first-order logic such as FO(ID), or an ASP language,
or a modelling language from the CP community such as ESSENCE \cite{ESSENCE}.

Recall that a vocabulary is a set of non-logical (predicate and function)
symbols. An interpretation for a vocabulary is provided by a {\em structure},
which consists of a set, called the domain or universe and denoted by $dom(.)$,
together with a collection of relations and (total) functions over the universe.
A structure can be viewed as an {\em assignment} to the elements of
the vocabulary. An expansion of a structure $\cal A$ is a structure $\cal B$
with the same universe, and which has all the relations and functions of
$\cal A$, plus some additional relations or functions.
The task of model expansion for an arbitrary logic $\cal L$ (abbreviated
$\cal L$-MX), is:
\begin{tabbing}
xx\=xx\=xxx.xxx\= \kill
\> \textbf{Model Expansion for logic $\cal L$}\\
\>\> \underline{Given:} \> {\sl 1.}\ An $\cal L$-formula $\phi$ with vocabulary
  $\sigma \cup \varepsilon$\\
\>\> \> {\sl 2.}\ A structure $\cal A$ for $\sigma$\\
\>\> \underline{Find:} an expansion of $\cal A$, to $\sigma \cup \varepsilon$,
   that satisfies $\phi$.
\end{tabbing}

\ignore{
\underline{Given:}   An $\cal L$-formula $\phi$ with vocabulary 
  $\sigma \cup \varepsilon$;
and a structure $\cal A$ for $\sigma$, \underline{Find:} an expansion of $\cal A$, to $\sigma \cup \varepsilon$,  that satisfies $\phi$.
}

Thus, we expand the structure $\cal A$ with relations and functions to interpret
$\varepsilon$, obtaining a model \cB{} of $\phi$. 
We call $\sigma$, the
vocabulary of $\cal A$, the {\em instance} vocabulary, and
$\varepsilon := vocab(\phi)\setminus \sigma$ the {\em expansion} vocabulary\footnote{By ``$:=$'' we mean ``is by definition'' or ``denotes''.}. 
%

\begin{example}\label{Ex:colouring}
The following formula $\phi$ of first order logic constitutes an MX
specification for Graph 3-colouring:

\begin{small}
$$
\begin{array}{c}
\forall x\ [ (R(x) \vee B(x) \vee G(x)) \\ 
\land \neg ( (R(x) \wedge B(x)) \lor (R(x) \land G(x)) \lor (B(x) \land G(x)))]\\
\wedge \ \forall x \forall y\  [E(x,y) \supset ( \neg (R(x) \wedge R(y)) \\
\ \ \ \ \ \ \ \ \land 
 \neg ( B(x) \wedge B(y)) \wedge 
 \neg ( G(x) \wedge G(y)))].
\end{array}
$$
\end{small}

An instance is a structure for vocabulary $\sigma = \{E\}$, i.e., a graph
${\cal A} = {\cal G} = ( V; E)$. The task is to find an interpretation for the
symbols of the expansion vocabulary $\varepsilon =\{ R, B, G\}$ such that the
expansion of $\cA$ with these is a model of $\phi$:
\begin{small}

$$
\underbrace{\overbrace{(V; E^{\cal A}}^{\cal A}, \ 
R^{\cal B}, B^{\cal B}, G^{\cal B} )}_{\cal B} \models\phi.
$$
\end{small}
The interpretations of $\varepsilon$, for structures $\cal B$ that satisfy
$\phi$, are exactly the proper 3-colourings of $\cal G$.
\end{example}

Given a specification, we can talk about a set of
$\sigma \cup \varepsilon$-structures which satisfy the specification.
Alternatively, we can simply talk about a set of
$\sigma \cup \varepsilon$-structures as an MX-task, without mentioning a
particular specification the structures satisfy. This abstract view makes our
study of modularity language-independent.


\subsection{Modular Systems}
\label{sec:modular}

This section reviews the concept of a modular system defined in \cite{TT:FROCOS:2011-long}
based on the initial development in \cite{JOJN}. As in \cite{TT:FROCOS:2011-long},
{\em each modular system abstractly represents an MX task}, i.e., a set (or class) of structures over
some instance and expansion vocabulary. 
A modular system is formally described as a set of primitive modules (individual MX tasks)
combined using the operations of: 
\begin{compactenum}
\item Projection($\pi_\tau(M)$) which restricts the vocabulary of a module,
\item Composition($M_1 \rhd M_2$) which connects outputs of $M_1$ to inputs of $M_2$,
\item Union($M_1 \cup M_2$), 
\item Feedback($M[R=S]$) which connects output $S$ of $M$ to its inputs $R$ and,
\item Intersection($M_1 \cap M_2$).
\end{compactenum}

Formal definitions of these operations are not essential for understanding
this paper, thus, we refer the reader to \cite{TT:FROCOS:2011-long} for details. 
The algebraic operations are illustrated in Examples \ref{ex:timetabling} and
\ref{ex:smt-solver}. In this paper, we only consider modular systems which do
not use the union operator.

{\em Our goal in this paper is
to solve the  MX task for a given modular system, i.e., given a modular
system $M$ (described in algebraic terms using the operations above) and
structure $\cA$, find a structure $\cB$ in $M$ expanding $\cA$. } We
find our inspiration in existing solver architectures by viewing them at
a high level of abstraction.

\begin{figure}
  \centering
    \includegraphics[scale=0.7]{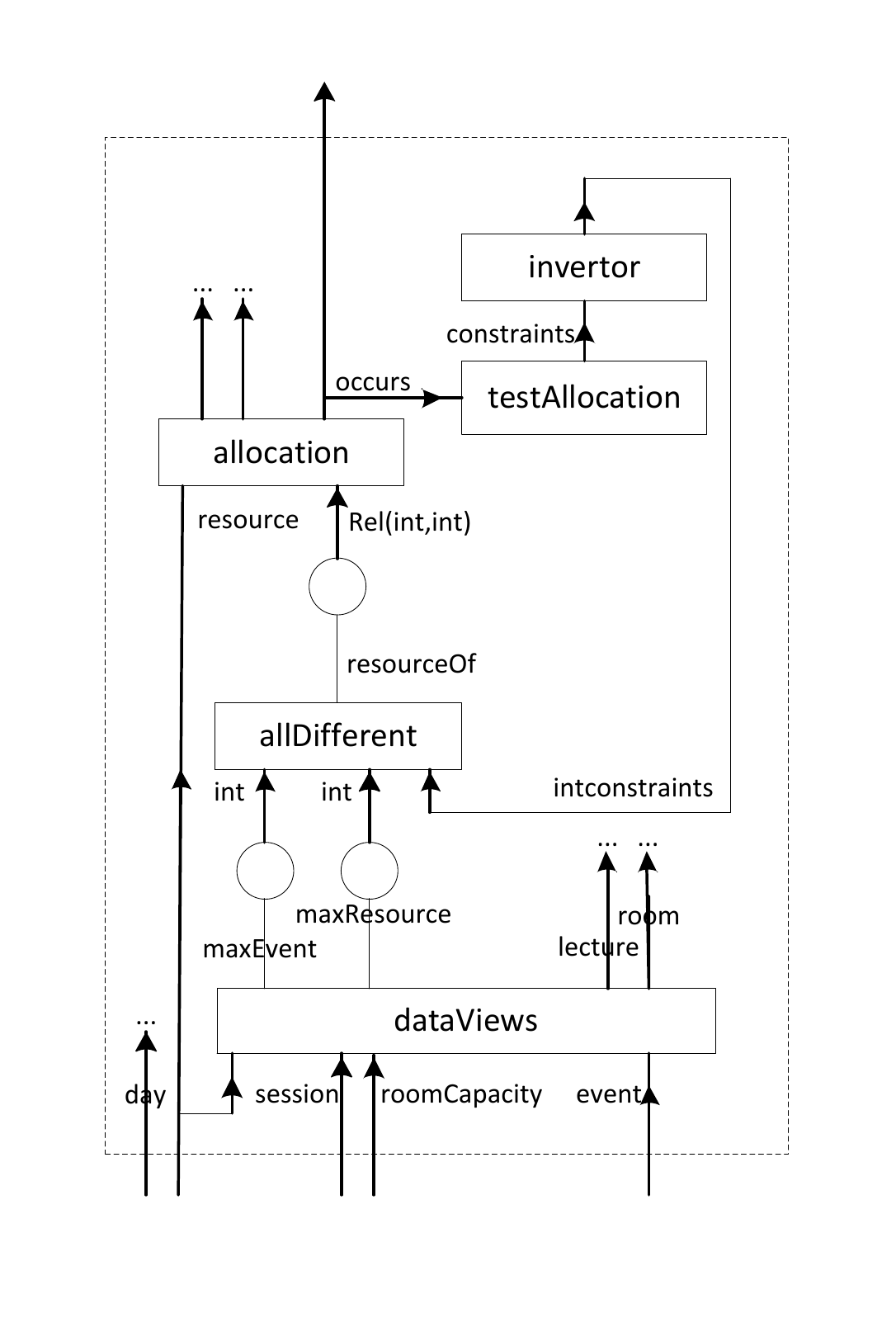}
  \caption{Modular System Representing a Timetabling Problem}
\label{fig:scheduling}
\end{figure}

\begin{example}[Timetabling \cite{JOJN}]\label{ex:timetabling}
Here, we use the example of timetabling from \cite{JOJN} and modify its
representation using our additional feedback operator. Figure \ref{fig:scheduling}
shows the new modular representation of the timetabling problem where the event
data and the resource data are the inputs and a list of events with their
associated sessions and resources (locations) is the output. This timetabling is
done so that the allocations of resource and sessions to the events do not
conflict. Unlike \cite{JOJN} where the ``allDifferent'' module is completely
independent of the ``testAllocation" module, here, through our feedback operator,
these modules are inter-dependent. This inter-dependency provides a better model
of the whole system by making the model closer to the reality. Also, here,
unlike \cite{JOJN} module ``allDifferent'' can be a deterministic module. In
fact, as proved in \cite{TT:FROCOS:2011-long}, the non-determinacy of all NP
problems can be modeled through the feedback operator. As will be shown later in
this paper, the existence of such loops can also help us to speed up the solving
process of  some problems.

In this paper, we propose an algorithm such that: given a modular system as in
Figure \ref{fig:scheduling} and given the inputs to this modular system, the
algorithm finds a solution to (or a model of) the given modular system, i.e., an
interpretation to the symbol ``occurs'' on this example that is not in conflict
with the constraints on the timetable.
\end{example}

\begin{SCfigure}\label{fig:smt-solver}
  \centering
    \includegraphics[scale=0.4]{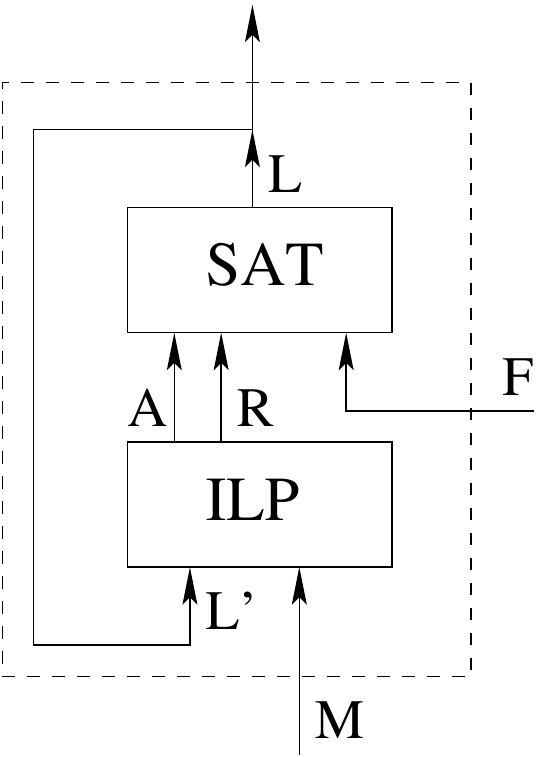}
  \caption{Modular System Representing an SMT Solver for the Theory of Integer Linear
    Arithmetic}
\end{SCfigure}

\begin{example}[SMT Solvers]\label{ex:smt-solver}
Consider Figure \ref{fig:smt-solver}: The inner boxes (with solid borders) show
simpler MX modules and the outer box shows our module of interest. The
vocabulary consists of all symbols $A$, $R$, $L$, $L'$, $M$ and $F$ where $A$,
$R$ and $L'$ are internal to the module, and others form its interface. Also,
there is a feedback from $L$ to $L'$.

Overall, this modular system describes a simple SMT solver for the theory of
Integer Linear Arithmetic ($T_{ILA}$). Our two MX modules are SAT and ILP.
They work on different parts of a specification. The ILP module takes a set
$L'$ of literals and a mapping $M$ from atoms to linear arithmetic formulas. It
returns two sets $R$ and $A$. Semantically, $R$ represents a set of subsets of
$L'$ so that $T_{ILA} \cup M|_r$\footnote{For a set $\tau$ of literals, $M|_\tau$
denotes a set of linear arithmetical formulas containing: (1) $M$'s image of
positive atoms in $\tau$, and (2) the negation of $M$'s image of negative atoms
in $\tau$.} is {\em unsatisfiable} for all subsets $r \in R$. Set $A$ represents
a set of propagated literals together with their justifications, i.e., a set of
pairs $(l, Q)$ where $l$ is an unassigned literal (i.e., neither $l \in L'$ nor
$\neg l \in L'$) and $Q$ is a set of assigned literals asserting $l \in L'$,
i.e., $Q \subseteq L'$ and $T_{ILA} \cup M|_Q \models M|_l$ (the ILA formula
$M|_l$ is a logical consequence of ILA formulas $M|_Q$). The SAT module takes
$R$ and $A$ and a propositional formula $F$ and returns set $L$ of literals such
that: (1) $L$ makes $F$ true, (2) $L$ is not a superset of any $r \in R$ and,
(3) $L$ respects all propagations $(l,Q)$ in $A$, i.e., if $Q \subseteq L$ then
$l \in L$. Using these modules and our operators, module SMT is defined as below
to represent our simple SMT solver:

\begin{equation}\label{formula:SMT}
SMT := \pi_{\{F,M,L\}}((ILP \rhd SAT)[L=L']).
\end{equation}

The combined module SMT is correct because, semantically, $L$ satisfies $F$ and all models in it
should have $R=\emptyset$, i.e., $T_{ILA} \cup M|_L$ is satisfiable. This is
because ILP contains structures for which if $r \in R$, then $r \subseteq L'=L$.
Also, for structures in SAT, if $r \in R$ then $r \not \subseteq L$. Thus, to
satisfy both these conditions, $R$ has to be empty. Also, one can easily see
that all sets $L$ which satisfy $F$ and make $T_{ILA} \cup M|_{L}$ satisfiable
are solutions to this modular system (set $A=R=\emptyset$ and $L'=L$).

So, there is a one-to-one correspondence between models of the modular system
above and SMT's solutions to the propositional part. To find a solution, one can
compute a model of this modular system. Note that, looking at modules as operators,
all models of module SMT are its fixpoints.
\end{example}
A description of a modular system  (\ref{formula:SMT})
looks like a formula in some logic.
One can define a satisfaction relation for that logic, however it is not needed here.
Still, since each modular system is a set of structures, we call the structures in a
modular system  {\em models} of that system. We are looking for models of a
modular system $M$ which expand a given instance structure $\cal A$. We call
them {\em solutions of $M$ for $\cal A$}.

\section{Computing Models of Modular Systems}
\label{sec:algorithm}
In this section, we introduce an algorithm which takes a modular system $M$ and
a structure $\cA$ and finds an expansion $\cB$ of $\cA$ in $M$. Our algorithm
uses a tool external to the modular system (a solver). It uses modules of a modular
system to ``assist'' the solver in finding a model (if one exists). Starting from an
empty expansion of $\cA$ (i.e., a partial structure which contains no information
about the expansion predicates), the solver gradually extends the current
structure (through an interaction with the modules of the given modular system)
until it either finds a model that satisfies the modular system or concludes
that none exists. To model this procedure, a definition of a partial structure
is needed.

\subsection{Partial Structures}

Recall that a structure is a domain together with an interpretation of a vocabulary. A
partial structure, however, may contain unknown values. For example, for a structure $\cB$ and
a unary relation $R$, we may know that $\langle 0 \rangle \in R^{\cB}$ and $\langle 1 \rangle \not \in R^{\cB}$,
but we may not know whether $\langle 2 \rangle \in R^{\cB}$ or $\langle 2 \rangle \not \in R^{\cB}$. Partial structures
deal with gradual accumulation of knowledge.

\begin{definition}[Partial Structure]\label{partial_structure}
We say $\cB$ is a $\tau_p$-partial structure over vocabulary $\tau$ if:
\begin{compactenum}
\item $\tau_p \subseteq \tau$,
\item $\cB$ gives a total interpretation to symbols in $\tau \backslash \tau_p$ and,
\item for each $n$-ary symbol $R$ in $\tau_p$, $\cB$ interprets $R$ using two sets $R^+$
and $R^-$ such that $R^+ \cap R^-= \emptyset$, and $R^+ \cup R^- \subsetneq (dom(\cB))^n$. 
\end{compactenum}

We say that $\tau_p$ is the partial vocabulary of $\cB$. If $\tau_p = \emptyset$,
then we say $\cB$ is total. For two partial structures $\cB$ and $\cB'$ over the
same vocabulary and domain, we say that $\cB'$ {\em extends} $\cB$ if all
unknowns in $\cB'$ are also unknowns in $\cB$, i.e., $\cB'$ has at least as much
information as $\cB$.
\end{definition}

If a partial structure $\cB$ has enough information to satisfy or falsify a formula $\phi$,
then we say $\cB \models \phi$, or $\cB \models \lnot \phi$, respectively.
Note that for partial structures, $\cB \models \lnot \phi$ and $\cB \not\models \phi$
may be different. We call a $\varepsilon$-partial structure $\cB$ over $\sigma \cup \varepsilon$
the {\em empty expansion} of $\sigma$-structure $\cA$, if $\cB$ agrees with
$\cA$ over $\sigma$ but $R^+=R^-=\emptyset$ for all $R \in \varepsilon$.

In the following, by structure we always mean a total structure, unless otherwise specified.
We may talk about ``bad'' partial structures which, informally, are the ones that
cannot be extended to a structure in $M$. Also, when we talk about a
$\tau_p$-partial structure, in the MX context, $\tau_p$ is always a subset of $\varepsilon$.

Total structures are partial structures with no unknown values. Thus, in the
algorithmic sense, total structures need no further guessing and should only be
checked against the modular system. A good algorithm rejects ``bad'' partial
structures sooner, i.e., the sooner a ``bad'' partial structure is detected, the faster
the algorithm is.

Up to now, we defined partial and total structures and talked about modules
rejecting ``bad'' partial structures. However, modules are sets of structures
(in contrast with sets of partial structures). Thus, acceptance of a partial
structure has to be defined properly. Towards this goal, we first formalize the
informal concept of ``good'' partial structures.
The actual acceptance procedure for partial structures is defined later in the section.

\begin{definition}[Good Partial Structures]\label{def:good-partial}
For a set of structures $S$ and partial structure $\cB$, we say $\cB$ is a
{\em good partial structure wrt $S$} if there is $\cB' \in S$ which extends $\cB$.
\end{definition}

\subsection{Requirements on the Modules}

Untill now, we have the concept of partial structures that the solver can work on,
but, clearly, as the solver does not have any information about the internals of the
modules, it needs to be assisted by the modules. Therefore, the next question could
be: ``what assistance does the solver need from modules so that its correctness is
always guaranteed?'' Intuitively, modules should be able to tell whether the solver
is on the ``right'' direction or not, i.e., whether the current partial structure is bad, and
if so, tell the solver to stop developing this direction further. We accomplish this
goal by letting a module accept or reject a partial structure produced by the solver
and, in the case of rejection, provide a ``reason'' to prevent the solver from producing
the same model later on. Furthermore, a module may ``know'' some extra information
that solver does not. Due to this fact, modules may give the solver some hints to
accelerate the computation in the current direction. Our algorithm models such hints
using ``advices'' to the solver.

Note that reasons and advices we are now talking about are different from predicate
symbols $R$ and $A$ in Example \ref{ex:smt-solver}. While, conceptually, $R$ and
$A$ also represent reasons and advices there, to our algorithm, they are just predicate
symbols for which an interpretation has to be found. On the other hand, reasons and
advices used by our algorithm are not specific to a modular system. They are entities
known to our algorithm which contain information to guide the solver in its search for a
model.

Also note that in order to pass a reason or an advice to a solver, there should
be a common language that the solver and the modules understand (although it may
be different from all internal languages of the modules). We expect this
language to have its own model theory and to support basic syntax such as
conditionals or negations. We expect the model theory of this language to 1) be
monotone: adding a sentence can not decrease the set of consequences and 2) have
resolution theorem which is the converse of the deduction theorem, i.e., if
$\Gamma \models A \supset B$ then $\Gamma \cup \{A\} \models B$. The presence of
the resolution theorem guarantees that, once an advice of form $Pre \supset Post$
is added to the solver, and when the solver has deduced $Pre$ under some
assumptions, it can also deduce $Post$ under the same assumptions. From now on,
we assume that our advices and reasons are expressed in such a language.

We talked about modules assisting the solver, but a module is a set of structures and
has no computational power. Instead, we associate each module with an ``oracle'' to
accept/reject a partial structure and give ``reasons'' and ``advices'' accordingly. Note
that it is unreasonable to require a strong acceptance condition from oracles because,
for example, assuming access to oracles which accept a partial structure iff it is a
good partial structure, one can always find a total model by polynomially many queries
to such oracles. While theoretically possible, in practice, access to such oracles is
usually not provided. Thus, we have to (carefully) relax our assumptions for a weaker
procedure (what we call a Valid Acceptance Procedure).

\begin{definition}[Advice]
Let $Pre$ and $Post$ be formulas in the common language of advices and reasons,
Formula $\phi:=Pre\supset Post$ is an {\em advice} wrt a partial structure $\cB$ and 
a set of structures $M$ if:
\begin{compactenum}
  \item $\cB\models Pre$,
  \item $\cB \not\models Post$ and,
  \item for every total structure $\cB'$ in $M$, we have $\cB'\models \phi$.
\end{compactenum}
\end{definition}
The role of an advice is to prune the search and to accelerate extending a partial
structure $\cB$ by giving a formula that is not yet satisfied by $\cB$, but is always
satisfied by any total extensions of $\cB$ in $M$. $Pre$ corresponds to the part that is
satisfied by $\cB$ and $Post$ corresponds to the unknown part that is not yet
satisfied by $\cB$.

\begin{definition}[Valid Acceptance Procedure]
Let $S$ be a set of $\tau$-structures. We say that $P$ is a
valid acceptance procedure for $S$ if for all $\tau_p$-partial structures
$\cB$, we have:
\begin{compactitem}
  \item If $\cB$ is total, then
    if $\cB \in S$, then $P$ accepts $\cB$,
    and if $\cB \not\in S$, then $P$ rejects $\cB$.
  \item If $\cB$ is not total but $\cB$ is good wrt $S$, then $P$ accepts $\cB$.
  \item If $\cB$ is neither total nor good wrt $\cB$, then $P$ is free to either accept or reject $\cB$.
\end{compactitem}
\end{definition}

The procedure above is called valid as it never rejects any good partial structures.
However, it is a weak acceptance procedure because it may accept some bad
partial structures. This kind of weak acceptance procedures are abundant in
practice, e.g., Unit Propagation in SAT, Arc-Consistency Checks in CP, and
computation of Founded and Unfounded Sets in ASP. As these examples show,
such weak notions of acceptance can usually be implemented efficiently as they
only look for local inconsistencies. Informally, oracles accept/reject a partial
structure through a valid acceptance procedure for a set containing all possible
instances of a problem and their solutions. We call this set a Certificate Set.

In theoretical computer science, a problem is a subset of $\{0,1\}^*$.
In logic, a problem corresponds to a set of structures. Here, we use the logic notion.

\begin{definition}[Certificate Set]
Let $\sigma$ and $\varepsilon$ be instance and expansion vocabularies. Let
$\cal P$ be a problem, i.e., a set of $\sigma$-structures, and $C$ be
a set of $(\sigma \cup \varepsilon)$-structures. Then, $C$ is a {\em
$(\sigma \cup \varepsilon)$-certificate set for $\cal P$} if for all $\sigma$-structures
$\cA$: $\cA \in {\cal P}$ iff there is a structure $\cB \in C$ that expands $\cA$.
\end{definition}

Oracles are the interfaces between our algorithm and our modules. Next we present
conditions that oracles should satisfy so that their corresponding modules can
contribute to our algorithm.

\begin{definition}[Oracle Properties]\label{def:cav-oracle}
Let $\cal L$ be a formalism with our desired properties. Let $\cal P$ be a problem, and
let $O$ be an oracle. We say that $O$ is
\begin{compactitem}
\item {\em Complete and Constructive (CC) wrt $\cal L$} if $O$ returns a reason
$\psi_\cB$ in $\cal L$ for each partial structure $\cB$ that it rejects such that: (1)
$\cB \models \lnot \psi_\cB$ and, (2) all total structures accepted by $O$ satisfy $\psi_\cB$.

\item {\em Advising (A) wrt $\cal L$} if $O$ provides a set of advices in $\cal L$
wrt $\cB$ for all partial structures $\cB$.

\item {\em Verifying (V)} if $O$ is a valid acceptance procedure for some certificate
set $C$ for $P$.
\end{compactitem}
\end{definition}

Oracle $O$ is {\em complete wrt $\cal L$} because it ensures the existence of
such a sentence and {\em constructive} because it provides such a sentence.
Oracle $O$ differs from the usual oracles in the sense that it does not only give
yes/no answers, but also provides reasons for why the answer is correct. It is {\em
advising} because it provides some facts that were previously unknown to guide the
search. Finally, it is {\em verifying} because it guides the partial structure to a
solution through a valid acceptance procedure. Although the procedure can be weak
as described above, good partial structures are never rejected and $O$ always
accepts or rejects total structures correctly. This property guarantees the
convergence to a total model. In the following sections, we use the term CCAV
oracle to denote an oracle which is complete, constructive, advising, and verifying.

\subsection {Requirements on the Solver}
In this section, we discuss properties that a solver needs to satisfy. Although the
solver can be realized by many practical systems, for them to
work in an orderly fashion and for algorithm to converge to a solution fast,
it has to satisfy certain properties. First, the solver has to be online
since the oracles keep adding reasons and advices to it. Furthermore, to guarantee
the termination, the solver has to guarantee progress, which means it either reports
a proper extension of the previous partial structure or, if not, the solver is guaranteed to
never return any extension of that previous partial structure later on. Now, we give the requirements on the
solver formally.

\begin{definition}[Complete Online Solver]
A solver $S$ is {\em complete and online} if the following conditions
are satisfied by $S$:
\begin{compactitem}
  \item $S$ supports the actions of initialization, adding sentences, and
reporting its state as either $\langle UNSAT \rangle$ or $\langle SAT, \cB \rangle$.
  \item If $S$ reports $\langle UNSAT \rangle$ then the set of sentences added
to $S$ are unsatisfiable,
  \item If $S$ reports $\langle SAT, \cB \rangle$ then $\cB$ does not falsify
any of the sentences added to $S$,
  \item If $S$ has reported $\langle SAT, \cB_1 \rangle, \cdots, \langle SAT, \cB_n \rangle$
and $1 \leq i < j \leq n$, then either $\cB_j$ is a proper extension of $\cB_i$ or,
for all $k \geq j$, $\cB_k$ does not extend $\cB_i$.
\end{compactitem}
\end{definition}

A solver as above is guaranteed to be sound (it returns partial structures that
at least do not falsify any of the constraints) and complete (it reports
unsatisfiability only when unsatisfiability is detected and not when, for
example, some heuristic has failed to find an answer or some time limit is
reached). Also, for finite structures, such a solver guarantees that our algorithm
either reports unsatisfiability or finds a solution to modular system $M$ and
instance structure $\cA$.

\subsection{Lazy Model Expansion Algorithm}

In this section, we present an iterative algorithm to solve model expansion 
tasks for modular systems. Algorithm \ref{alg:lazy-mx} takes an instance structure
and a modular system (and its CCAV oracles) and integrates them with a complete
online solver to iteratively solve a model expansion task. The algorithm works by
accumulating reasons and advices from oracles and gradually converging to a
solution to the problem.

The role of the reasons is to prevent some bad structures and their extensions from being
proposed more than once, i.e., when a model is deducted to be bad by an oracle, a new
reason is provided by the oracle and added to the solver such that all models of the system
satisfy that reason but the ``bad'' structure does not. The role of an advice is to
provide useful information to the solver (satisfied by all models) but not yet satisfied by
partial structure $\cB$. Informally, an advice is in form ``if Pre then Post'', where ``Pre''
corresponds to something already satisfied by current partial structure $\cB$ and ``Post''
is something that is always satisfied by all models of the modular system satisfying the
``Pre'' part, but not yet satisfied by partial structure $\cB$. It essentially tells the solver
that ``Post'' part is satisfied by all intended structures (models of the system) extending
$\cB$, thus helping the solver to accelerate its computation in its current direction.

The role of the solver is to provide a possibly good partial structure to the oracles,
and if none of the oracles rejects the partial structure,
keep extending it until we find a solution or conclude none exists.
If the partial structure is rejected by any one of the oracles, the solver gets a reason
from the oracle for the rejection and tries some other partial structures.
The solver also gets advices from oracles to accelerate the search.

\ignore{
The role of the solver is to take a formula $\phi$ and a partial structure $\cB$
and to do one of the following: (1) to return an extension $\cB'$ of $\cB$
such that $\cB' \not\models \lnot \phi$, (2) to announce that $\phi$ is unsatisfiable, or (3) to
announce that no extension of $\cB$ satisfies $\phi$.

During this process, constraints are accumulated in three different ways: (1) a
(possibly partial) structure is rejected by some oracle (associated to a module)
and a new constraint is given by the oracle, (2) a set of advises is given by some oracle, 
or (3) the solver decides that none
of the extensions of a particular structure are promising and gives a constraint
to prevent us from following that direction.
}

\begin{algorithm}\label{alg:lazy-mx}
\KwData{Modular System $M$ with each module $M_i$ associated with a CCAV oracle $O_i$, input structure $\cA$ and complete online solver $S$}
\KwResult{Structure $\cB$ that expands $\cA$ and is in $M$}
  \Begin{
    Initialize the solver $S$ using the empty expansion of $\cA$ \;
    \While{TRUE}
    {
      Let $R$ be the state of $S$ \;
      \lIf {$R=\langle UNSAT \rangle$}
      {
        \Return{Unsatisfiable} \;
      }
      \ElseIf {$R=\langle SAT, \cB \rangle$}
      {
        Add the set of advices from oracles wrt $\cB$ to $S$ \;
        \If {$M$ does not accept $\cB$}
        {
          Find a module $M_i$ in $M$ such that $M_i$ does not accept $\cB |_{vocab(M_i)}$ \;
          Let $\psi$ be the reason given by oracle $O_i$ \;
          Add $\psi$ to $S$ \;
        }
        \lElseIf {$\cB$ is total}
        {
          \Return{$\cB$} \;
        }
      }
    }
  }
\caption{Lazy Model Expansion Algorithm}
\end{algorithm}

\section{Examples: Modelling Existing Frameworks}
\label{sec:modelling}
In this section, we describe algorithms from three different areas and show that
they can be effectively modelled by our proposed algorithm in the context of model
expansion. Note that our purpose here is not to analyze other systems but to show
the effectiveness of our algorithm in the absence of an implementation. We establish
this claim by showing that our algorithm acts similar to the state-of-the-art
algorithms when the right components are provided.

\subsection{Modelling DPLL($T$)}\label{sec:dpll_t}

DPLL($T$) \cite{DPLL_T} system is an abstract framework to model the lazy SMT approach.
It is based on a general DPLL($X$) engine, where $X$ can be instantiated with a theory $T$ solver.
DPLL($T$) engine extends the Decide, UnitPropagate, Backjump, Fail and Restart
actions of the classic DPLL framework with three new actions: (1) {\bf TheoryPropagate} gives
literals that are $T$-consequences of current partial assignment, (2) {\bf $T$-Learn} learns
$T$-consistent clauses, and (3) {\bf $T$-Forget} forgets some previous lemmas of theory solver.

To participate in DPLL($T$) solving architecture, a theory solver provides three operations:
(1) taking literals that have been set true, (2) checking if setting these literals true is $T$-consistent
and, if not, providing a subset of them that causes inconsistency, (3) identifying some currently
undefined literals that are $T$-consequences of current partial assignment and providing a
justification for each. More details can be found in \cite{DPLL_T}.

The modular system $DPLL(T)$ of the DPLL($T$) system is the same as the one in Example
\ref{ex:smt-solver}, except that we have module $M_P$ instead of $SAT$ and $M_T$ instead of $ILP$.
In Figure \ref{fig:smt-solver}, $A$ corresponds to the result of TheoryPropagate action that
contains some information about currently undefined values in $L'$ together with their justifications;
$R$ is calculated from the $T$-Learn action and corresponds to reasons of $M_T$ rejecting $L'$.
\ignore{
\here{Where to put???
$T$-Forget
is modelled inside the solver $S$, i.e., $S$ needs not use all constraints passed to it to generate an
extended partial model. It may only use constraints that have relatively been active in its DPLL-based
solving procedure and forget about the rest temporarily \here{not quite right, T-forget means forget it forever}. 
Finally, Fail corresponds to the case of algorithm returning ``Unsatisfiable''. }
}

To model DPLL($T$), we introduce a solver $S$ to be any DPLL-based online SAT
solver, so that it performs the basic actions of Decide, UnitPropagate, Fail,
Restart, and also Backjump when the backjumping clause is added the solver. The
two modules $M_T$ and $M_P$ are attached with oracles $O_T$ and $O_P$
respectively. They accept a partial structure $\cB$ iff their respective module
constraints is not falsified by $\cB$. When rejecting $\cB$, a reason
``$P_ {in}$ then $P_ {out}$'' (true about all models of the module) is returned
where $P_ {in}$ (resp. $P_ {out}$) is a property about input (resp. output)
vocabulary of the module satisfied (resp. falsified) by $\cB$. They may also
return advices of the same form but with $P_ {out}$ being neither satisfied nor
falsified by $\cB$. The constructions of these two modules are similar; so, we
only give a construction for the solver $S$ and module $M_T$:

{\bf Solver $S$} is a DPLL-based online SAT solver (clearly complete and online).

{\bf Module $M_T$} 
The associated oracle $O_T$ accepts a partial structure $\cB$ if it does not
falsify the constraints described in Example \ref{ex:smt-solver} on $L'$, $M$, $A$, and $R$ for module $M_T$.
If $\cB$ is rejected, $O_T$ returns a reason $\psi := \psi_{in} \supset \psi_{out}$ where
$\cB|_ {\{L',M\}} \models \psi_{in}$ but $\cB|_ {\{A,R\}} \models \neg \psi_ {out}$.
Clearly, $\cB \models \lnot \psi$ and all models in $M_T$ satisfy $\psi$.
Thus, $O_T$ is complete and constructive. 
$O_T$ may also return some advices which are similar to the reason above except that $\psi_{out}$
is neither satisfied nor falsified by $\cB$. Hence, $O_T$ is an advising oracle.
Also, $O_T$ always makes the correct decision for a total structure and rejects a partial structure only
when it falsifies the constraints for $M_T$. $O_T$ never rejects any
good partial structure $\cB$ (although it may accept some bad non-total structures). Therefore, $O_T$ is
a valid acceptance procedure for $M_T$ and, thus, a verifying oracle.

\begin{proposition}
1. Modular system $DPLL(T)$ models the DPLL($T$) system.
2. Solver $S$ is complete and online.
3. $O_P$ and $O_T$ are CCAV oracles.
\end{proposition}

\ignore{
\here{The proof of correctness of above proposition is omitted, since the correctness can be easily seen from the fact that DPLL-based SAT solvers are complete and constructive.}
}

DPLL(T) architecture is known to be very efficient and many solvers are designed to use it, including
most SMT solvers \cite{SMT}. The DPLL(Agg) module \cite{DPLL_AGG} is suitable for all DPLL-based
SAT, SMT and ASP solvers to check satisfiability of aggregate expressions in DPLL($T$) context. All
these systems are representable in our modular framework.

\subsection{Modelling ILP Solvers}\label{sec:ilp}
Integer Linear Programming solvers solve optimization problems. In this paper, we model ILP
solvers which use general branch-and-cut method to solve {\em search} problems instead, i.e.,
when target function is constant. We show that Algorithm \ref{alg:lazy-mx} models such ILP solvers. 
ILP solvers with other methods and Mixed Integer Linear Programming solvers use similar
architectures and, thus, can be modelled similarly. 

The search version of general branch-and-cut algorithm \cite{ILP} is as follows: 
\begin{compactenum}
  \item Initialization: $S=\{$ILP$^0\}$ with ILP$^0$ the initial problem.
  \item Termination: If $S=\emptyset$, return UNSAT.
  \item Problem Select: Select and remove problem ILP$^i$ from $S$.
  \item Relaxation: Solve LP relaxation of ILP$^i$ (as a search problem). If infeasible, go to
step 2. Otherwise, if solution $X^{iR}$ of LP relaxation is integral, return solution $X^{iR}$.
  \item Add Cutting Planes: Add a cutting plane violating $X^{iR}$ to relaxation and go
to 4.
  \item Partitioning: Find partition $\{C^{ij}\}^{j=k}_{j=1}$ of constraint set $C^i$ of problem ILP$^i$.
Create $k$ subproblems ILP$^{ij}$ for $j=1,\cdots,k$, by restricting the feasible region of
subproblem $ILP^{ij}$ to $C^{ij}$. Add those $k$ problems to $S$ and go to step 2. Often, in practice,
finding a partition is simplified by picking a variable $x_i$ with non-integral value $v_i$ in $X^{iR}$
and returning partition
$\{C^i \cup \{x_i\leq \lfloor v_i \rfloor\}, C^i \cup \{x_i\geq \lceil v_i \rceil\}\}$.
\end{compactenum}

\begin{SCfigure}\label{fig:ilp-solver}
  \centering
  \includegraphics[scale=0.6]{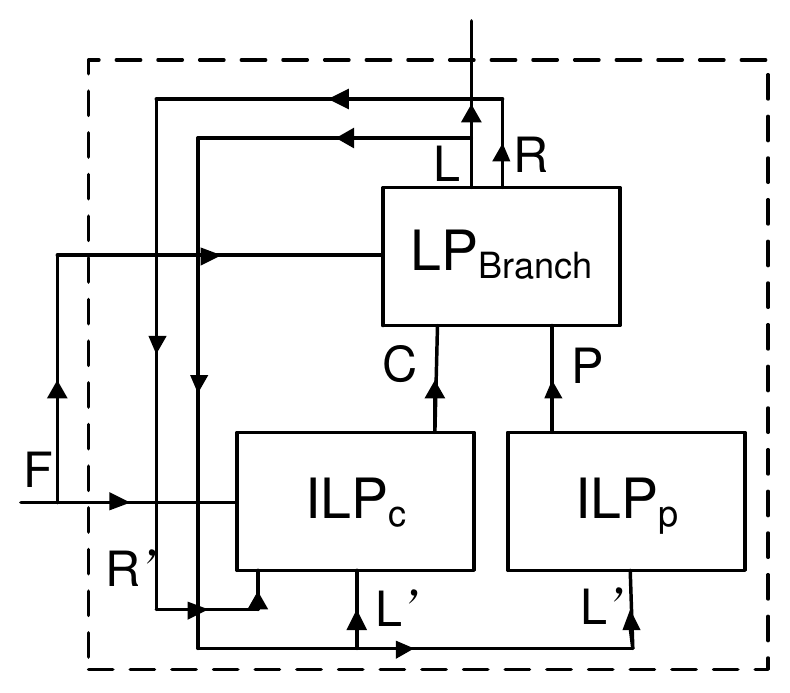}
  \caption{Modular System Representing an ILP Solver}
\end{SCfigure}

We use the modular system shown in figure \ref{fig:ilp-solver} to represent the ILP solver.
The $ILP_c$ module takes the problem specification $F$, a set of assignments $L'$
and a set of range information $R'$ (which, in theory, describes assumptions about ranges of
variables and, in practice, describes the branch information of an LP+Branch solver)
as inputs and returns a set $C$.
When all the assignments in $L'$ are integral, $C$ is empty, and if not,
$C$ represents a set of cutting planes conditioned by a subset of range information $R'$,
i.e., set of linear constraints that are violated by $L'$ and all the set of assignments
satisfy both $F$ and $R'$ also satisfy the set of
cutting planes. 
The $ILP_p$ module only takes the set of assignments $L'$ as input and outputs a set of partitioning clauses
$P$, such that when all the assignments in $L'$ is integral, $P$ is empty and when
there is a non-integral assignment to a variable $x$, $P$ is a set of partitioning clauses indicating that
assignment to x should be either less than or equal to $\lfloor L'(x) \rfloor$ or greater than
or equal to $\lceil L'(x) \rceil$. The $LP_{Branch}$ module takes $F$, $C$ and $P$ as inputs and
outputs the set of assignment $L$ and the set of range information $R$ such that $L$ satisfies specification $F$,
the range information $R$, the set of conditional cutting planes in $C$, and the set of partitioning clauses in $P$.
We define the compound module $ILP$ to be:

$$
ILP := \pi_{\{F,L\}}(((ILP_c \cap ILP_p) \rhd LP_{Branch})[L=L'][R=R']).
$$

The module $ILP$ defined above is correct because all models satisfying it should have
$C=B=\emptyset$ because, $ILP_c$ contains structures in which, for every $(S,c) \in C$,
which denotes cutting plane $c$ under condition $S$, we have $S \subseteq R'=R$ and
$c$ is violated by $L'=L$. Furthermore, $LP_{Branch}$ contains structures in which $L$
satisfies both $R$ and $C$, which indicates that either $S$ is
not the subset of $R$ or $L$ satisfies $c$. Thus $C$ is empty. By a similar argument, one can
prove that $B$ also has to be empty.

We compute a model of this modular system by introducing a solver that interacts with all of
the three modules above. We introduce an extended LP solver $S$ which allows us to deal
with disjunctions. $S$ performs a depth-first-like search on disjunctive constraints, and runs its
internal LP solver on non-disjunctive constraints plus the range information the search branch corresponds
to. So, Partitioning action of ILP corresponds to adding a disjunctive constraint to the solver.
All three modules above are associated with oracles $O_c$, $O_p$ and $O_{lp}$, respectively.
Exact constructions are similar to the ones in section \ref{sec:dpll_t}.
Here we only give a construction for the solver $S$:

\ignore{
which accept a partial structure $\cB$ if it does
not falsify the constraints on relations in each module and in the case of the rejection,
return a reason in the form of ``$P_{in}$ then $P_{out}$'' where $P_{in}$ and $P_{out}$ are
properties about, respectively, input and output of the module and $\cB$ satisfies $P_{in}$
and falsifies $P_{out}$ (while it should not). They may also return some advices of the same
form such that $\cB$ satisfies $P_{in}$ but neither satisfies nor falsifies $P_{out}$.
The construction of these three modules is similar; so we only give a construction for solver
$S$ and module $ILP_c$:
}

{\bf Solver $S$} is an extended LP solver, i.e., uses an internal LP solver. Let $Br$ denote the set of
branch constraints in $S$ (constraints involving disjunctions) and $L$ denote the set of pure
linear constraints. When new constraint is added to $S$, $S$ adds it to $Br$ or $L$ accordingly.
$S$ then performs a depth-first-like search on branch clauses, and, at branch $i$, passes
$L \cup Br_i$ (a set of linear constraints specifying branch $i$) to its internal LP solver. 
Then, $S$ returns $\langle SAT, \cB \rangle$ if LP finds a model $\cB$ in some branch,
and $\langle UNSAT \rangle$ otherwise.
Note that, by construction, $S$ is complete and online.

\ignore{
{\bf Module $ILP_c$:} The associated oracle $O_c$ accepts a partial structure $\cB$ if it does not
falsify the constraints described above on $F$, $R'$, $L'$, and $C$ for $ILP_c$ module.
If $\cB$ is rejected, $O_c$ returns a reason $\psi := \psi_{in} \supset \psi_{out}$ where
$\psi_{in}$ is satisfied by $F^{\cB}\cup L'^{\cB}\cup R'^{\cB}$ but $\psi_{out}$ is falsified by $C^{\cB}$. 
Clearly, $\cB \models \lnot \psi$ and all models $\cB' \in ILP_c$ satisfy $\psi$.
Thus, $O_c$ is complete and constructive. 
$O_c$ may also return some advices which are similar to the reason above except that it requires $C^{\cB}$
to be neither satisfied nor falsified by $\cB$. It is not hard to see that $O_c$ is also an advising oracle.
Finally, $O_c$ always makes the correct decision for a total structure and rejects a partial structure only
when it falsifies the constraints for the module. Thus, $O_c$ never rejects any
good partial structure $\cB$ (although it may accept some bad non-total structures). Therefore, $O_c$ is
a valid acceptance procedures for $ILP_c$ and, thus, a verifying oracle.
}

\begin{proposition}
1. Modular system $ILP$ models the branch-and-cut-based ILP solver.
2. $S$ is complete and online.
3. $O_c$, $O_p$ and $O_{lp}$ are CCAV oracles. 
\end{proposition}

\ignore{Correctness of proposition \ref{ILP:S} can be easily seen from the 
completeness and constructiveness of the LP solvers and the construction of the solver $S$.}

There are many other solvers in ILP community that use some ILP or MILP solver as their low-level
solver. It is not hard to observe that most of them also have similar architectures that can be closely
mapped to our algorithm.

\subsection{Modelling Constraint Answer Set Solvers}\label{sec:casp}
The Answer Set Programming (ASP) community puts a lot of effort into optimizing
their solvers. One such effort addresses ASP programs with variables ranging
over huge domains (for which, ASP solvers alone perform poorly due to the huge
memory the grounding uses). However, embedding Constraint Programming (CP)
techniques into ASP solving is proved useful because grounding such variables is
partially avoided.

In \cite{Baselice05towardsan}, the authors extend the language of ASP and its reasoning method
to avoid grounding of variables with large domains by using constraint solving techniques.
The algorithm uses ASP and CP solvers as black boxes and non-deterministically extends a
partial solution to the ASP part and checks it with the CP solver. Paper
\cite{Mellarkod:2008} presents another integration of answer set generation and
constraint solving in which a traditional DPLL-like backtracking algorithm is
used to embed the CP solver into the ASP solving.

Recently, the authors of \cite{ASP-CP-combination} developed an improved hybrid
solver which supports advanced backjumping and conflict-driven no good learning
(CDNL) techniques. They show that their solver's performance is comparable to
state-of-the-art SMT solvers. Paper \cite{ASP-CP-combination} applies a partial
grounding before running its algorithm, thus, it uses an algorithm on
propositional level. A brief description of this algorithm follows: Starting
from an empty set of assignments and nogoods, the algorithm gradually extends
the partial assignments by both unit propagation in ASP and constraint
propagation in CP. If a conflict occurs (during either unit propagation or
constraint propagation), a nogood containing the corresponding unique
implication point (UIP) is learnt and the algorithm backjumps to the decision
level of the UIP. Otherwise, the algorithm decides on the truth value of one of
the currently unassigned atoms and continues to apply the propagation. If the
assignment becomes total, the CP oracle queries to check whether this is
indeed a solution for the corresponding constraint satisfaction problem (CSP).
This step is necessary because simply performing constraint propagation on the
set of constraints, i.e., arc-consistency checking, is not sufficient to decide
the feasibility of constraints.

The modular model of this solver is very similar to the one in Figure
\ref{fig:smt-solver}, except that we have module $ASP$ instead of $SAT$ and $CP$
instead of $ILP$. The compound module $CASP$ is defined as:

$$
CASP := \pi_{\{F,M,L\}}((CP \rhd ASP)[L=L']).
$$

As a CDNL-like technique is also used in SMT solvers, the above algorithm is
modelled similarly to Section \ref{sec:dpll_t}. We define a solver $S$ to be a
CDNL-based ASP solver. We also define modules $ASP$ and $CP$ to deal with the
ASP part and the CP part. They are both associated oracles similar to those
described in Section \ref{sec:dpll_t}. We do not include the details here as
they are similar to the ones in section \ref{sec:dpll_t}.

\ignore{
\begin{proposition}
1. Modular system $CASP$ models the constraint answer set solver described above.
2. $S$ is complete and online.
3. The oracles are CCAV oracles.
\end{proposition}
}
Note that one can add reasons and advices to an ASP solver safely in the form of
conflict rules because stable model semantics is monotonic with respect to such
rules. Also, practical CP solvers do not provide
reasons for rejecting partial structures. This issue is dealt with in
\cite{ASP-CP-combination} by wrapping CP solvers with a conflict analysis
mechanism to compute nogoods based on the first UIP scheme.

\section{Extension: Approximations}
\label{sec:approx}
Almost all practical solvers use some kind of propagation technique. However,
in a modular system, propagation is not possible in general because nothing is
known in advance about a module. According to \cite{TT:FROCOS:2011-long}, it
turns out that knowing only some general information about modules such as their
totality and monotonicity or anti-monotonicity, one can hugely reduce the search
space.

Moreover, paper \cite{TT:FROCOS:2011-long} proposes two procedures to approximate models of
what are informally called positive and negative feedbacks. These procedures
correspond to least fixpoint and well-founded model computations (but in modular
setting). Here, we extend Algorithm \ref{alg:lazy-mx} using these procedures. The
extended algorithm prunes the search space of a model by propagating information
obtained by these approximation procedures to the solver. First, let us define some
properties that a module may satisfy.

\begin{definition}[Module Properties \cite{TT:FROCOS:2011-long}]
Let $M$ be a module and $\tau$, $\tau'$ and $\tau''$ be some subsets of $M$'s
vocabulary. $M$ is said to be:
\begin{compactenum}
  \item {\bf $\tau$-total over a class $C$ of structures} if by restricting
models of $M$ to vocabulary $\tau$ we can obtain all structures in $C$.
  \item {\bf $\tau$-$\tau'$-$\tau''$-monotone (resp.
$\tau$-$\tau'$-$\tau''$-anti-monotone)} if for all structures $\cB$ and $\cB'$
in $M$ we have that if $\cB|_{\tau} \sqsubseteq \cB'|_{\tau}$ and
$\cB|_{\tau'} = \cB'|_{\tau'}$ then $\cB|_{\tau''} \sqsubseteq \cB'|_{\tau''}$
(resp. $\cB'|_{\tau''} \sqsubseteq \cB|_{\tau''}$).
\end{compactenum}
\end{definition}

In \cite{TT:FROCOS:2011-long}, it is shown that these properties are fairly general
and that, given such properties about basic MX modules, one can derive
similar properties about complex modules.

Now, we can restate the two approximation procedures from \cite{TT:FROCOS:2011-long}.
For $\cB$ and $\cB'$ over the same domain, but distinct vocabularies,
let $\cB||\cB'$ denote the structure over that domain and $voc(\cB) \cup voc(\cB')$
where symbols in $voc(\cB)$ [resp. $voc(\cB')$] are interpreted as in $\cB$ [resp. $\cB'$].
We first consider the case of a positive feedback, i.e.,
when relation $R$ which increases monotonically when $S$ increases is fed back
into $S$ itself. This procedure is defined for a module $M':=M[S=R]$ and partial
structure $\cB$ where $M$ is ($\tau \cup \{S\}$)-total and $\{S\}$-$\tau$-$\{R\}$-monotone
and $\cB$ gives total interpretation to $\tau$. It defines a chain $L_i$ of interpretations for
$S$ as follows:
\begin{equation}\label{eq:pos-feedback}
\begin{array}{l}
L_0:=S^{+^\cB}, \\
L_{i+1}:=R^{M(\cB|_\tau~||~\cL)} \mbox{ where } dom(\cL)=dom(\cA) \mbox{ and } S^\cL=L_i.
\end{array}
\end{equation}

\begin{algorithm}\label{alg:lazy-mx-approximation}
\KwData{ Similar to Algorithm \ref{alg:lazy-mx}, but with modules' totality,
monotonicity and anti-monotonicity properties given}
\KwResult{Structure $\cB$ expands $\cA$ and is in $M$}
  \Begin{
    Initialize the solver $S$ using the empty expansion of $\cA$ \;
    \While{true}
    {
      Let $R$ be the state of $S$ \;
      \lIf {$R=\langle UNSAT \rangle$}
      {
        \Return{Unsatisfiable} \;
      }
      \ElseIf {$R=\langle SAT, \cB \rangle$}
      {
        Add the set of advices from oracles wrt $\cB$ to $S$ \;
        \If {$M$ does not accept $\cB$}
        {
          Find a module $M_i$ in $M$ such that $M_i$ does not accept $\cB |_{vocab(M_i)}$ \;
          Let $\psi$ be the reason given by oracle $O_i$ \;
          Add $\psi$ to $S$ \;
        }
        \lElseIf {$\cB$ is total}
        {
          \Return{$\cB$} \;
        }
        \Else
        {
          \ForEach{applicable positive feedback $M_1:=M_2[T=T']$}
          {
            Let $L^*$ be the limit of series $L_i$ in Equation \ref{eq:pos-feedback} \; 
            Propagate $L^* \backslash T^{+^\cB}$ to $S$ \;
          }
          \ForEach{applicable negative feedback $M_1:=M_2[T=T']$}
          {
            Let $\langle L^*, U^* \rangle$ be the limit of series $\langle L_i, U_i \rangle$ in Equation \ref{eq:neg-feedback} \;
            Propagate $(L^* \backslash T^{+^\cB}) \cup \mathbf{not}([dom(\cA)]^n \backslash (U^* \cup T^{-^\cB}))$ to $S$ \;
          }
        }
      }
    }
  }
\caption{Lazy Model Expansion with Approximation (Propagation)}
\end{algorithm}

The procedure for a negative feedback is similar, but for $M$ being
$\{S\}$-$\tau$-$\{R\}$-anti-monotone. It defines an increasing sequence $L_i$ and
a decreasing sequence $U_i$ which say what should be in added to $S^+$ and $S^-$.
Here, $n$ is the arity of relations $R$ and $S$:
\begin{equation}\label{eq:neg-feedback}
\begin{array}{c}
L_0 := S^+, U_0 := [dom(\cA)]^n \backslash S^-, \\
L_{i+1}:=R^{M(\cB|_\tau~||~\cU)} \mbox{ where } dom(\cU)=dom(\cA) \mbox{ and } S^\cU=U_i, \\
U_{i+1}:=R^{M(\cB|_\tau~||~\cL)} \mbox{ where } dom(\cL)=dom(\cA) \mbox{ and } S^\cL=L_i.
\end{array}
\end{equation}

Now, Algorithm \ref{alg:lazy-mx} can be extended to Algorithm
\ref{alg:lazy-mx-approximation} which uses the procedures above to find new
propagated information which has to be true under the current assumptions, i.e.,
the current partial structure. This information is sent back to the
solver to speed up the search process. Algorithm \ref{alg:lazy-mx-approximation}
needs a solver which can get propagated literals.

\section{Related Works}
\label{sec:rel-work}
This paper is a continuation of \cite{TT:FROCOS:2011-long} and proposes an
algorithm for solving model expansion tasks in the modular setting. The modular
framework of \cite{TT:FROCOS:2011-long} expands the idea of model theoretic (and
thus language independent) modelling of \cite{JOJN} and introduces the feedback
operator and discusses some of the consequences (such as complexity
implications) of this new operator. There are many other works on modularity in
declarative programming that we only briefly review.

An early work on adding modularity to logic programs is \cite{Eiter-1997}. The
authors derive a semantics for modular logic programs by viewing a logic program
as a generalized quantifier. This is further generalized in \cite{Tommi-modular-equiv}
by considering the concept of modules in declarative programming and introducing
modular equivalence in normal logic programs under the stable model semantics.
This line of work is continued in \cite{Tommi-dlp-modularity} to define
modularity for disjunctive logic programs. There are also other approaches to
adding modularity to ASP languages and ID-Logic as described in
\cite{Baral-2006,Balduccini-2007,DT-ToCL-2008:long}.

The works mentioned earlier focus on the theory of modularity in declarative
languages. However, there are also works that focus on the practice of modular
declarative programming and, in particular, solving. These works generally fall
into one of the two following categories:

The first category consists of practical modelling languages which incorporate
other modelling languages. For example, X-ASP \cite{XASP} and ASP-PROLOG
\cite{ASP-PROLOG} extend prolog with ASP. Also ESRA \cite{ESRA}, ESSENCE
\cite{ESSENCE} and Zinc \cite{Zinc} are CP languages extended with features from
other languages. However, these approaches give priority to the host language
while our modular setting gives equal weight to all modelling languages that are
involved. It is important to note that, even in the presence of this distinction,
such works have been very important in the development of this paper because
they provide guidelines on how a practical solver deals with efficiency issues.
We have emphasized on this point in Section \ref{sec:modelling}.

The second category consists of the works done on multi-context systems. In
\cite{Brewka-2007:long}, the authors introduce non-monotonic bridge rules to the
contextual reasoning and originated an interesting and active line of research
followed by many others for solving or explaining inconsistencies in
non-monotonic multi-context systems
\cite{Eiter:DMCS:long,Eiter:MCS:IE:1:long,Eiter:MCS:IE:2:long,Eiter:MCS:AIE:long}.
However, these works do not consider the model expansion task. Moreover, the
motivations of these works originate from distributed or partial knowledge, e.g.,
when agents interact or when trust or privacy issues are important. Despite
these differences, the field of multi-context systems is very relevant to our
research. Investigating this connection as well as incorporating results from
the research on multi-context system into our framework is our most important
future research direction.

\section{Conclusion}
\label{sec:Conclusion}
\ignore{
Each communities in the field of declarative problem solving
have their own ways of solving problems.
People from different communities normally do not share their expertise in designing solvers.
It is necessary to do some research on bridging previous works done in different communities
so that they can learn from each other to obtain more powerful solvers.

We took a language-independent view on iterative modular problem solving
and proposed an algorithm which actually solves the problems described in the modular systems
in the context of model expansion. We defined the conditions on
modules such that once satisfied, modules with possibly different languages
can participate in the algorithm. We argued that our algorithm captures the
essence of the practical solvers, by showing that DPLL($T$) framework, ILP
solvers and the state of the art combinations of ASP and CP are all
specializations of our modular framework.
}

We addressed the problem of finding  solutions to a modular system in the context of 
model expansion and proposed an algorithm which finds such solutions. 
We defined conditions on modules such that once satisfied, modules described  
with possibly different languages
can participate in our algorithm. We argued that our algorithm captures the
essence of  practical solvers, by showing that DPLL($T$) framework, ILP
solvers and state-of-the-art combinations of ASP and CP are all
specializations of our modular framework.
We believe that our work bridges  work done in different communities and 
contributes to cross-fertilization of the fields.

\ignore{
We also believe that our work can be used
to design and implement new solvers that integrate several languages. 
}

\bibliography{Files/fb}
\bibliographystyle{Files/splncs}

\end{document}